\pdfoutput=1
\documentclass[10pt,a4paper]{article}
\RequirePackage{latexsym,amsmath,amssymb}
\RequirePackage[dvipsnames,usenames]{color}

\usepackage{fullpage}

\usepackage[british]{babel}
\usepackage[latin1]{inputenc}
\usepackage[T1]{fontenc}
\usepackage[final]{showkeys} % [draft/final]
\usepackage{hyperref}
\hypersetup{pdfauthor={M. Cadoni, M. Tuveri...},
            pdftitle={}
            }
\usepackage[]{cleveref}
\Crefname{equation}{Eq.}{Eqs.}
\Crefname{figure}{Fig.}{Figs.}
\crefformat{plural}{#2eqs.~(#1)#3}
\crefname{section}{Sect.}{Sects.}

\usepackage{amsfonts}
\def\lb{\label}
\def\be#1\ee{\begin{align}#1\end{align}}
\newcommand{\ppar}{{p_\parallel}}

\newcommand{\mb}{m_{\rm B}}
\renewcommand{\le}{\leqslant}
\renewcommand{\leq}{\leqslant}

\def\lb{\label}

\title{\bf Galactic dynamics and long-range quantum gravity}
\author{M.~Cadoni${}^{ab}$\thanks{E-mail: mariano.cadoni@ca.infn.it},
and 
M.~Tuveri${}^{ab}$\thanks{E-mail: matteo.tuveri@ca.infn.it}
\\
\\
${}^a$\emph{Dipartimento di Fisica, Universit\`a di Cagliari}
\\
{\em Cittadella Universitaria, 09042 Monserrato, Italy}
\\
\\
${}^b$\emph{I.N.F.N, Sezione di Cagliari, Cittadella Universitaria, 09042 Monserrato, Italy}
\\
\\}
\begin{document}
\maketitle
\begin{abstract}
We explore in a systematic way the possibility that long-range quantum gravity effects  could 
play a role at galactic scales and could be responsible for the phenomenology commonly attributed 
to dark matter. We argue that the presence of baryonic matter breaks the scale symmetry of 
the de Sitter (dS) spacetime generating an IR scale $r_0$,  corresponding to the scale at 
which the typical dark matter effects we observe in galaxies arise. It  also generates a huge 
number of bosonic excitations with wavelength larger than the size of the cosmological horizon 
 and in thermal equilibrium with dS spacetime. 
We show that for $r\gtrsim r_0$ these excitations produce a new component for the radial 
acceleration of stars in galaxies which leads to the result found by McGaugh {\sl et al.} by 
fitting  a large amount of observational data and with the MOND theory.  We also propose 
a generalized thermal equivalence principle and use it to give another independent derivation of 
our result. 
Finally, we show that our result can be also derived as the weak field limit of Einstein's general 
relativity sourced by an anisotropic fluid.

\end{abstract}
%%%%%%%%%%%%%%%%%%%%%%%%%%%%%%%%%%%%%%%%%%%%%%
%%%%%%%%%%%%%%%%%%%%%%%%%%%%%%%%%%%%%%%%%%%%%%
% 						Introduction
%%%%%%%%%%%%%%%%%%%%%%%%%%%%%%%%%%%%%%%%%%%%%%
%%%%%%%%%%%%%%%%%%%%%%%%%%%%%%%%%%%%%%%%%%%%%%

\section{Introduction}
Quantum mechanics determines the macroscopic behaviour of many physical systems. 
This was realized since its discovery as it was used to explain macroscopic phenomena 
for which classical physics failed, like e.g. black body radiation and the heat capacity in solids.  
By now, we also know very well that  quantum macroscopic systems exist  as a 
result of long-range quantum coherent  states. We have several examples of this,  
including Bose-Einstein (BE) condensates, superfluids, neutron stars etc.
This is believed to be true for physical systems in which the electromagnetic (or strong) 
interaction plays the main role, but its extension to gravitational systems is commonly overlooked.
Indeed, due to the weakness of the gravitational interaction, quantum gravity effects are usually 
believed to be relevant only at very small scales of the order of Planck length $l_p=\sqrt{\hbar G}$  
\footnote{In this paper we use, unless explicitly stated,  units $c=k_B=1$, where $c$ is 
the speed of light, $k_B$ is the Boltzmann constant and $G$ denotes the Newton constant}.

In general, our description of gravitational phenomena is based on two implicit 
assumptions:  $1)$ quantum gravity corrections are suppressed by inverse powers of the
Planck mass, $m_p=\sqrt{\hbar /G}$; $2)$ effective field theories, 
give a reliable description of infrared (IR) physics.
Thus, quantum gravity is expected to play a role near a spacetime singularity but to be 
completely irrelevant at larger scales. As a consequence, the gravitational physics at 
galactic and solar system scales should be fully determined by general relativity (GR) or its 
eventual classical modifications. 

However, it is logically possible that the common wisdom expressed by the two statements 
above is not always true. Some of the laws governing long times and long distances  dynamics 
in our universe could have, in principle, a quantum mechanical origin. There are several reasons 
supporting this point of view like, for instance, the fact  that long-range quantum effects 
play an important role in several systems like e.g. Bose-Einstein  condensates. Their 
peculiar features are determined not by the strength of the interaction but, rather, by genuine 
quantum features like e.g. the Pauli exclusion principle or the existence of  quantum collective 
states.
In particular, condensed matter systems offer a nice  paradigm  for what concerns long-range 
quantum interactions. When applied to gravity  this  perspective suggests  that macroscopic 
quantum gravity effects 
could  be originated from  the cooperative result of a huge number of quantum modes of extremely 
large wavelength.

Recently, several proposal of  long-range quantum gravity effects have been put forward. 
The emergent gravity scenario, in its different  forms, considers spacetime and gravity as a 
macroscopic manifestation of microscopic quantum gravity degrees of freedom 
~\cite{Sakharov:1967pk,Jacobson:1995ab,Padmanabhan:2009vy,Padmanabhan:2014ypu,
Jacobson:2015hqa,Oriti:2016acw,Padmanabhan:2016eld,Verlinde:2016toy,Linnemann:2017hdo,
De:2019fva}. 
 For example, in the corpuscular gravity scenario black holes and the de Sitter universe 
may be explained as quantum coherent states of a large number of gravitons
~\cite{Dvali:2010bf,Dvali:2010ue,Dvali:2010jz,Dvali:2011th,Dvali:2011aa,Dvali:2012en,
Dvali:2012rt,Binetruy:2012kx,Mueck:2013mha,Casadio:2016zpl,Casadio:2017cdv,
Cadoni:2017evg,Cadoni:2018dnd}. 
 Recent investigations on the topic have also suggested that quantum gravity corrections 
may also be relevant at the horizon of astrophysical black holes
~\cite{Compere:2019ssx,Hooft:2019nmf,Blommaert:2019hjr}.

One of the most promising arena for looking for long-range quantum gravity effects is 
galactic dynamics. 
Despite its great successes in explaining  present experimental data about the accelerated 
expansion of the universe~\cite{Riess:1998cb}, structure formation, galaxy rotation curves 
and gravitational lensing, the $\Lambda$CDM model of standard cosmology~\cite{Penzias:1965wn, Ade:2013zuv} 
fails to explain the so-called baryonic Tully-Fisher relation~\cite{Tully:1977fu,McGaugh:2000sr}, 
$v^2\simeq\sqrt{a_0\, G\, \mb}$. This formula establishes a relation between
the asymptotic velocity $v$ of stars in galaxies  and the total baryonic mass $\mb$.  
 In particular, the acceleration parameter $a_0$ is of the same order of magnitude of 
the current value of the Hubble constant $H$, $a_0\simeq 1.2\times 10^{-10} m\ s^{-2}$~\cite{McGaugh:2016leg}.

At phenomenological level a simple explanation of the Tully-Fisher relation is offered by 
Milgrom's MOdified Newtonian Dynamics~(MOND)~\cite{Milgrom:1983ca,Milgrom:2014usa}. 
In the MOND theory, $a_0$ is promoted to a fundamental constant of nature. 
The modification of Newtonian gravity encoded in MOND can be explained either as 
a modification of the laws of inertia or as an additional acceleration component, $a_{\rm MOND}=\sqrt{a_0a_{\rm B}}$, 
where $a_B$ is the Newtonian acceleration due to baryonic matter. 
From astrophysical observations we know that these new 
effects arise when $a_B\simeq a_0$, to which corresponds a critical scale $r_0\simeq \sqrt{Gm_B/a_0}$.

Motivated by these considerations, recently, there have been some alternative proposal
to explain the galactic-scale phenomenology commonly attributed to dark matter. 
For example, in one of these approaches the additional MOND acceleration is associated 
to a "dark force" (DF) generated by the reaction of dark energy (DE) to the 
presence of baryonic matter~\cite{Verlinde:2016toy,Cadoni:2017evg,Cadoni:2018dnd,Hossenfelder:2017eoh,Dai:2017guq,Cai:2017asf}.
In an another approach, this is the result of an environmental modification of the 
inertial/gravitational mass ratio~\cite{Smolin:2017kkb}.
In both approaches the modification of the Newtonian laws of gravity is thought as the 
macroscopic manifestation of a long-range/late times quantum gravity effect
~\footnote{The idea that MOND can be an expression of quantum gravity has been 
considered also by  by other authors~\cite{Milgrom:1998sy,2014arXiv1411.2665V,
Woodard:2014wia,Modesto:2010rm,Ho:2011xc,Hendi:2010xr,
2016EL....11569001M,Klinkhamer:2011un,Pikhitsa:2010nd,Ho:2010ca,
Edmonds:2013hba,Edmonds:2016tio,Ng:2016qvh}.}.

However, these approaches suffer from a serious drawback: they are able to reproduce 
the asymptotic MOND formula but their predictions significantly differ from observations 
in the galactic region~\cite{Milgrom:2016huh,Lelli:2017sul,Pardo:2017jun,Hees:2017uyk}. 
In particular,  in the MOND framework, the observational data about rotational curves 
of galaxies can be reproduced only by introducing a phenomenological function $F$.
This function interpolates between the  baryonic acceleration due to standard 
Newtonian gravity near to the galactic core and the (deep) MOND regime in the outer 
galactic region. 
An interpolating function $F$ (whose explicit value will be shown in section~\ref{subsect:rotcurves}
fitting a large amount of observational data coming from 
galaxies with different shapes (spiral, elliptical, spherical) has been proposed by 
McGaugh {\sl et al.}~\cite{McGaugh:2016leg,Lelli:2017vgz} 
(see however~\cite{Salucci:2018eie,DiPaolo:2018mae}).
\medskip

The purpose of this paper is to explore in a systematic way the possibility that long-range 
quantum gravity effects could play a relevant role at galactic scales, derive the macroscopic 
manifestation of these effects and compare them with observations. 
We will use simple and general arguments not relaying on any specific microscopic 
theory of gravity, but only on general features of thermodynamics, quantum and 
statistical mechanics and, finally, general relativity.

We will start in Sect.~\ref{sec1} by showing that the presence of baryonic matter breaks 
the scale isometry of the dS spacetime generating, in this way, a length-scale $r_0$ 
and  related long-range quantum gravity effects. The gravitational nature of 
 this long-range interaction implies the existence of a huge number of soft, 
spin-2,  collective bosonic excitations following a thermal Bose-Einstein distribution at the 
temperature of the dS horizon.  
In Sect.~\ref{sect2}, using a simple one-dimensional quantum mechanical model, 
we will compute the energy of these  collective bosonic excitations and the related IR scale $r_0$.  
In Sect.~\ref{sect:rot}, using simple thermodynamical arguments, we will derive the 
acceleration of a test mass produced at galactic scales by these DE bosonic 
excitations and compare our result with galactic observations. We will show that 
our theoretical result reproduces the observational data and the McGaugh interpolating 
function~\cite{McGaugh:2016leg,Lelli:2017vgz} up to a numerical proportionality factor 
between the acceleration parameter $a_0$ and the cosmological acceleration $H$. 
In sect. \ref{sec:4}, building up the results of Ref.~(\cite{Smolin:2017kkb}), we will propose a 
generalized thermal equivalence principle. We will use it to give another independent 
derivation of our results and to compute the acceleration proportionality factor, i.e. $1/2\pi$, 
in agreement with observational data~\cite{McGaugh:2016leg,Lelli:2017vgz}.
In sect. \ref{sect7} we will deal with  our collective  bosonic excitations in the context of 
corpuscular gravity. 
In sect. \ref{sec:8} we will show that our weak field description allows for a general 
covariant uplifting of the theory in terms of GR sourced by an anisotropic fluid. 
Finally in Sect. \ref{sec:9} we will present our conclusions.

%

%%%%%%%%%%%%%%%%%%%%%%%%%%%%%%%%%%%%%%%%%%%%%%%%%%%%%%
%%%%%%%%%%%%%%%%%%%%%%%%%%%%%%%%%%%%%%%%%%%%%%%%%%%%%%
%					QUANTUM UNIVERSE
%%%%%%%%%%%%%%%%%%%%%%%%%%%%%%%%%%%%%%%%%%%%%%%%%%%%%%
%%%%%%%%%%%%%%%%%%%%%%%%%%%%%%%%%%%%%%%%%%%%%%%%%%%%%%
%
\section{Long-range quantum gravity effects}
\lb{sec1}
To begin with, we consider our universe as made only by  dark energy  and baryons, 
i.e.  we do not assume the  presence of any exotic form  of matter like  dark matter.
Since we do not know what DE really is, we model it in the simplest way as a 
positive cosmological constant $\Lambda$.
In absence of baryonic matter, consistently with GR, our universe is described 
by a de Sitter (dS) spacetime with  cosmological horizon size  (the dS radius) 
$L=\sqrt{3/\Lambda}$. 
The cosmological acceleration $H$ is related to $L$ by $H=1/L$. 
The cosmological horizon has an associated Hawking temperature given by
%whereas $\lp$, $\mpl=1/\lp$ and 
%$G=\lp^2$ are the Planck length, the Planck mass and the Newton 
%constant, respectively.}. 
\be\lb{ht}
T_{dS}=\frac{\hbar}{2\pi L}.
\ee
Four dimensional dS spacetime can be defined has an hyperbola embedded in 
$\mathbb{R}^{1,4}$. In the static patch the empty, i.e. without baryonic matter,  
dS universe is described by the following metric
\be\lb{dSmetric}
ds^2=-f(r)dt^2+f(r)^{-1}dr^2+r^2d\Omega^2,\quad f(r)=1-\frac{r^2}{L^2},
\ee
where $d\Omega^2$ is the metric of the $2-$sphere.
The dS spacetime has an isometry group $SO(1,4)$ inherited from its embedding in 
$\mathbb{R}^{1,4}$ which acts as conformal transformations of the $3-$sphere at infinity. 
In particular, it has an intrinsic scale invariance, which becomes evident by writing the metric 
in FLRW form and in the flat slicing 
\be\lb{dSmetric1}
ds^2=-dt^2+e^{\frac{2t}{L}}d\vec{y}\,^2,
\ee
where $d\vec{y}\,^2$ is the flat metric on $\mathbb{R}^{3}$. 
The metric is invariant under $d\vec{y}\to \lambda d\vec{y}, t\to t- L\log \lambda$. 
The scale invariance means that the dS spacetime cannot be endowed with a length scale.
Every comoving length interval can be stretched during the cosmological expansion without 
changing the form of the metric~\eqref{dSmetric1}. 
In particular, this implies that the dS radius $L$ does not play any dynamical role rather, 
from a semi-classical perspective, it has to be thought just as a thermal 
scale whose inverse gives the (horizon) dS temperature (\ref{ht})~\cite{Narnhofer:1996zk,Deser:1997ri,Jacobson:1997ux}.  
 
Scale invariance is not a peculiarity of the dS background but a general property of the 
Einstein-Hilbert action with a cosmological constant term. In fact,  it has been shown that Einstein 
general relativity with cosmological constant, but in absence of matter, can be formulated as a 
scale-free theory \cite{Cadoni:2006ww}. 

Let us now introduce the baryonic matter in the dS universe. 
For simplicity, in this paper we consider baryonic matter in the form of a point-like mass $m_B$, 
but our computations can be easily generalized to the case of a spherically symmetric mass 
distribution $m_B(r)$.
Due to the presence of baryonic matter, the solution of Einstein equations of GR will change 
accordingly and they acquire a new, Schwarzschild-like, term in the metric function $f(r)$ in Eq.~(\ref{dSmetric}): 
$f(r)= 1-\frac{r^2}{L^2} -\frac{2G m_B}{r}$. 
Hence, the presence of the baryonic matter explicitly breaks the scale symmetry of the dS spacetime 
and a length scale $r_0$ is generated.  
As discussed in Ref. \cite{Cadoni:2006ww}, the breaking of the  scale symmetry due to the presence 
of  matter occurs also at the full Einstein-Hilbert  action level.

At first sight one could be led to identify $r_0$ with the 
Schwarzschild radius associated to $m_B$, $r_0= R_s=2Gm_B$. However, this is correct 
only if one assumes that DE  does not react to the presence of baryonic matter. 
The existence of  such a DE reaction is quite natural in an 
emergent gravity~\cite{Verlinde:2016toy} or corpuscular gravity~\cite{Cadoni:2017evg,Cadoni:2018dnd,Giusti:2019wdx,Hossenfelder:2017eoh,Dai:2017guq,Cai:2017asf} 
scenario and in view of our lack of understanding of DE represents the most plausible assumption. 
In this paper we will assume that this is the case. 

If DE reacts to the presence of baryonic matter the length scale $r_0$ can in principle depend on 
three scales: the Planck length $l_p$, which sets the microscopic scale of quantum gravity phenomena, 
$R_s$  and $L$. 
Using an analogy borrowed from condensed matter physics one can describe DE as a  system of size $L$ 
characterized by a microscopic scale $l_p\ll L$ (i.e. the dimension of its elementary constituents) in which 
we introduce an impurity of size $R_s$. Generically, one expects the generation of a {\sl mesoscopic} scale 
$r_0$ with $l_p\ll r_0<L$, which does not depend on $l_p$ and is completely determined by $R_s$ and $L$.

We will  determine $r_0$ shortly, meanwhile lets us ask the question about the relevance of quantum effects 
at the mesoscopic scale $r_0$. The question can be answered by comparing the Compton wavelength 
$\lambda_c$  associated to the impurity with $r_0$.
When we assume that DE does not react to baryonic matter we have seen that $r_0$ must be of the order 
of $R_s$ and $\lambda_c=\hbar/m_B$. 
As it is well known, in this case the condition $\lambda_c\simeq r_0$ determines the typical scale of quantum 
gravity effects to be of the order of Planck length $l_p\simeq 10^{-35} m$. We have reached the usual result 
that quantum gravity effects are completely negligible at macroscopic scales.

The situation in which DE reacts to the presence of baryonic matter is completely different.  In this case,
we expect the scale $r_0$ to depend also on $L$ and to become {\sl mesoscopic}. Moreover, in this picture, 
the quantum modes of DE can acquire new collective (quantum) gravitational properties, which become 
relevant at scales of order of $r_0$. The baryonic matter can be seen as an impurity in the dS system whose 
effects are not localised within its Schwarzschild radius, rather diffuse in a broad region (the galactic region).

To evaluate the Compton wavelength associated to the baryonic mass we must consider a test mass 
$m$ at distance $r$ from $m_B$ and its classical Newtonian energy, 
\be\lb{ne}
V_N(r)= - \frac{G mm_B}{r}.
\ee
The typical scale of quantum gravity effects is determined by
\be\lb{lc}
\lambda_c \simeq \frac{\hbar}{|V_N|}\simeq r_0.
\ee
From  Eqs~(\ref{ne}) and~(\ref{lc}) one finds that quantum effects are negligible 
for  $r\ll r_0R_s m/\hbar$ and become relevant for $r\simeq r_0 R_s m/\hbar$. 
We see that at scales  $r=r_0$  quantum effects become relevant when  $m=\hbar/R_s$, 
i.e. when the test mass becomes of the same order of the Compton mass of a black hole 
with mass $m_B$. In particular we can have quantum gravity effects at galactic scales if 
$r_0$ is of the order of magnitude of galactic radii.
This is a quite interesting result suggesting  that we can have long-range quantum 
gravity effects at large scales  when the test particle has a mass of the order 
of the Compton mass of a black hole built with the corresponding baryonic mass $m_B$. 
Notice that we can write $m\simeq m_p^2/m_B$ so that  $m\ll m_p$ if $m_B$ is big enough.

Having established that it is possible to have quantum gravity effects at large (galactic) scales 
generated by the presence of baryonic matter, we are faced with the problem of identifying 
the corresponding collective quantum excitation of the DE system.     
 Since we do not have a consistent quantum gravity description of the dS universe (see e.g. \cite{Witten:2001kn}), 
we can only formulate some general guess about the nature of the quantum excitation 
generated in the DE system by the presence of baryonic matter.  

Evidences from GR and various quantum gravity models indicate that  the
gravitational interaction should be mediated by bosonic spin-2 particles.  
This is, for instance, quite natural in a corpuscular gravity scenario,
like that considered  in Ref.~\cite{Cadoni:2018dnd} where the bosonic 
excitations have been identified as "dark gravitons". 
 This suggests that DE excitations can be considered as bosonic 
quantum states in thermal equilibrium with the dS spacetime described by 
a Bose-Einstein distribution with zero chemical potential at temperature $T_{dS}$ 
given by Eq.~(\ref{ht})
\be\lb{N_BE}
N(\varepsilon)=\frac{1}{e^{\frac{\varepsilon}{T_{dS}}}-1},
\ee
where $\varepsilon$ is the energy of the excitation. In the following we will refer to 
them as {\sl DE bosonic excitations}. Notice that we are assuming that the energy spectrum 
is non-degenerate.

It is well known that bosons can condense to produce a Bose-Einstein condensate (BEC) 
when the temperature  of the system drops below a certain critical temperature $T_c$. 
In our case we expect this critical temperature to be of the order of  $T_{dS}$.
The dS universe can be therefore considered as a quantum critical state representing a BEC. 
The description of the dS spacetime as a BEC also emerges quite naturally in a corpuscular 
gravity context~\cite{Binetruy:2012kx,Cadoni:2017evg,Cadoni:2018dnd}, where it has been 
considered as a BEC of a large number $N\sim L^2/l_p^2$ of gravitons with energy 
$\varepsilon\sim \hbar/L$.

It is interesting to notice that a quite similar phenomena happens at scales smaller than 
the galactic one when one considers black holes.
The same relations we have written before for the dS universe also hold for black holes 
 if we replace $L$ with the Schwarzschild radius $R_s$.
Black holes can be also described as a BEC constituted by a large number $N_{BH}\sim 
R_s^2/l_p^2$  of gravitons with typical energy  $\varepsilon_{B}\sim \hbar/R_s$
~\cite{Dvali:2010bf,Dvali:2010ue,Dvali:2010jz,Dvali:2011th,Dvali:2011aa,Dvali:2012en,
Dvali:2013eja,Dvali:2012rt,Binetruy:2012kx,Mueck:2013mha,Casadio:2016zpl,Casadio:2017cdv}.
Similarly to the case of the dS universe, also black holes are critical BEC. 
  
We note that our quantum description of the dS universe and of black holes is quite similar 
to that of typical Bose-Einstein condensates as superfluids (e.g. helium under a critical 
temperature) or superconductors, where the DE bosonic excitations play the role of phonons. 
The use of Bose-Einstein condensates in cosmology has been recently proposed also in~\cite{Das:2014agf,Das:2018udn,Das:2018lwl}.
We stress the fact that in the case of dS universe, the analogy with condensed matter systems 
and in particular with superconductors allows us to consider the baryonic matter as an impurity.
\medskip 

What is commonly believed is that quantum mechanics 
do not play any role at intermediate scales, i.e.~at galactic and extra-galactic scales, being 
important only at Planckian scales. 
However, if the length scale $r_0$  at which these new quantum effects arise is of the order 
of magnitude of the galaxy size, it is possible that many phenomena we observe at galactic 
scale could be associated to long-range quantum interaction between dark energy and 
baryonic matter. 
In this case, from the BE distribution (\ref{N_BE}) we see that the hard excitations 
with $\varepsilon\gg T_{dS}$ are exponentially suppressed. Hence the scale $r_0$ 
corresponds to the appearance of DE bosonic excitations, i.e.~$N={\cal O}(1)$, and 
the energy of the excitation must satisfy
\be\lb{gh}
\frac{\varepsilon}{T_{dS}}={\cal O}(1).
\ee 
As a consequence, we are left with a large number $N\gg 1$ of soft excitations 
with $\varepsilon\ll T_{dS}$. In this limit, at leading order we have
\be\lb{lim}
N(\varepsilon)=\frac{T_{dS}}{\varepsilon}.
\ee
Hence, we may generically expect the phenomenology we observe at galactic scale to be the macroscopic 
manifestation at scales $r>r_0$ of a huge number of these soft DE bosonic excitations, which are
associated to a long-range quantum interaction between dark energy and baryonic matter. 
Conversely, at scales $l_p\ll r\ll r_0$ these quantum effects are suppressed and the gravitational 
interaction is well described by the Newtonian and, more generally, general relativity theories. 

In order to determine $r_0$ we need to determine the energy $\varepsilon$ of DE bosonic excitations. 
Owing to its quantum gravitational origin a simple-minded guess for $\varepsilon$ could be 
a simple Compton form $\varepsilon\sim \hbar/r$.
However, due to the peculiar features of the interaction between DE and baryonic matter, 
the energy of the DE bosonic excitations could in principle deviate from this simple form. 
In general we can parametrize this new "dark" interaction term by introducing a 
dimensionless coupling constant $\alpha$ characterizing the interaction 
between the cosmological condensate and baryonic matter. We write  
\be\label{nf}
\varepsilon= \frac{\hbar\alpha}{r},
\ee
The coupling constant $\alpha$ will depend on both the 
baryonic coupling constant $Gm_B$ and the properties of the cosmological 
condensate,~i.e. the dS radius $L$.
Moreover, being the DE bosonic excitations soft we must require $\alpha<1$.

In the next subsection we determine the value of $\varepsilon$, i.e. of the 
dimensionless coupling constant $\alpha$,  and the value of the scale 
$r_0$ by considering a simple one-dimensional quantum model of 
galactic dynamics.

%%%%%%%%%%%%%%%%%%%%%%%%%%%%%%%%%%%%%%%%%%%%%%%%%%%%%%
\subsection {One-dimensional quantum mechanical model for galactic dynamics}
\lb{sect2}
In the region $r>r_0$ there is a new interaction term between the baryonic mass 
$m_B$ and the test particle of mass $m$ located at distance $r$, which has not 
 classical, but rather  quantum mechanical origin.
We can think of the test particle as dressed by the presence of dark energy 
and we want to derive the energy spectrum of the system.
Because we do not have any clear understanding of DE microphysics, we use again 
an analogy with condensed matter physics.

We generically expect the DE excitations introduced in the previous subsection to be 
collective modes of  DE, i.e.~quasi particles with effective mass $m^*$ 
subject to an effective potential $V(x)$. 
Owing to the spherical symmetry of the problem, we can use a one-dimensional 
quantum mechanical model described by a single coordinate $x$.
 The analogy with condensed matter systems becomes more and more stringent 
when we try to describe, at least at a mesoscopic level, the relation between DE, dS 
spacetime and the emergent properties of spacetime itself.  
Indeed, old and new emergent gravity scenarios~\cite{Sakharov:1967pk,Jacobson:1995ab,
Padmanabhan:2014ypu,Jacobson:2015hqa,Oriti:2016acw,Verlinde:2016toy,Linnemann:2017hdo,
De:2019fva} describe spacetime as a sort of medium where the internal degrees of freedom 
are located in specific points as they constitute a sort of cosmological lattice. 
In this way, if spacetime emerges from the dynamics of these microscopic degrees of 
freedom, it is quite natural to assign to it some elastic properties and, as a consequence, 
to have the generation of collective modes as quasi-particles, analogous to phonons. 
For these reasons, assuming that DE has some sort of elastic properties and working 
in the harmonic approximation, the presence of the baryonic mass $m_B$ will generate 
an elastic response  of the DE medium and, consequently, the harmonic oscillator 
effective potential,
\be\lb{harmos}
V(x)= \frac{1}{2} K(r,L)\, x^2.
\ee
The elastic constant $K$ in general will depend both from the distance $r$ from the baryonic 
mass and from the horizon radius $L$ of the dS spacetime.  
The simplest choice for $K$ is to take it proportional to the absolute value of the  Newtonian potential 
energy $|V_N(r)|$ of the mass $m^*$ in the potential generated by the  baryonic mass $m_B$. The 
proportionality factor must have dimensions of length$^{-2}$ defining a pivotal distance $\langle x\rangle$ 
such that $V(x=\langle x\rangle)=|V_N(r)|$. This gives
\be\lb{K}
K=\frac{2 Gm_Bm^*}{r\langle x\rangle^2}.
\ee
The pivotal distance $\langle x\rangle$ must keep information about the two length scales $r$ and $L$ 
and  in view of the difference of many order of magnitude between $r$ and $L$, it is appropriate 
to take a geometric mean of the two scales, $\langle x\rangle= \sqrt{r L}$. Hence, the elastic constant 
takes the following form
\be\lb{K1}
K=\frac{2 Gm_Bm^*}{r^2 L}.
\ee
The energy spectrum of the DE excitations is now  simply given by the energy levels $\varepsilon_n$ of the 
one-dimensional quantum harmonic oscillator of mass $m^*$ with potential given by Eq.~(\ref{harmos}) 
and elastic constant~(\ref{K1}).  We find
\be\lb{K1a}
\varepsilon_n=\left(n+\frac{1}{2}\right)\frac{\hbar}{r} \sqrt{\frac{2 Gm_B}{L}},
\ee
where the ground state energy is 
\be\lb{K2}
\varepsilon_0=\frac{\hbar}{r} \sqrt{\frac{Gm_B}{2L}}.
\ee
For a given baryonic mass $m_B$ and distance $r$ the low-lying excitations dominate.   
Neglecting a factor of order $1$ in Eq.~\eqref{K1a} the energy $\varepsilon$ of DE bosonic 
excitations is given by 
\be\label{oi}
\varepsilon\simeq\sqrt\frac{G m_B}{ L}\frac{\hbar}{r}.
\ee 
This equation shows that, as expected, the energy of DE bosonic excitations only depends on $Gm_B,L$ 
and $r$. The value of the coupling constant $\alpha$ in Eq.~\eqref{nf} is $\alpha\simeq\sqrt{G m_B/ L}$.

It is interesting to note that $\alpha$ is independent from the microscopic quantum gravity 
scale $l_p$, which can enter in $\alpha$ only trough the Newtonian coupling $Gm_B$. 
This means that the long-range quantum mechanical properties of dS universe manifest themselves at mesoscopic 
scales represented by $r_0$ and they are different from the quantum gravitational effects one expect 
at very small scales. 
As we will see later in this paper, $r_0$ represents the scale at which Einstein gravity 
has to be modified in order to fully describe galactic gravitational dynamics.

We are now  able to determine the length scale $r_0$. 
As already mentioned in the previous section, the presence of the baryonic mass 
will produce in the DE  condensate a large number $N$ of DE bosonic excitations 
of energy $\varepsilon$ in thermal equilibrium with dS universe at temperature 
$T_{dS}=\hbar/2\pi L$, distributed according to the Bose-Einstein statistics  
given by Eq.~\eqref{N_BE}.
The scale $r_0$ corresponds to the appearance of this excitation.
By using Eq.~(\ref{oi}) into Eq.  (\ref{gh}) one finds
\be\lb{ppp}
r_0\simeq\sqrt{R_s L}=\sqrt{G m_B L},
\ee
telling us that the mesoscopic scale $r_0$ corresponds, again, to the geometric mean 
between the Schwarzschild radius of the baryonic source and the cosmological scale $L$.
We can now express the coupling constant $\alpha$ of Eq. (\ref{nf}) in terms of $r_0$ and $L$ 
only, i.e.
\be\lb{hh}
\alpha= \frac{r_0}{L}.
\ee
The coupling constant $\alpha$ is suppressed by the inverse of the dS radius $L$, it is 
therefore extremely small unless $r_0$ becomes of the order of magnitude of galactic scales. 
This explains why longe-range quantum gravity effects become relevant only (at least) at 
galactic scales.
On the other hand the number $N$ of DE excitations is exponentially suppressed for $\varepsilon> T_{dS}$, 
which in view of Eqs.~(\ref{oi}) and~(\ref{ppp}) implies $r<r_0$, whereas $N$ grows according 
to Eq. (\ref{lim}) for $\varepsilon\ll T_{dS}$. 
 This again shows that long-range quantum gravity effects become relevant only at galactic scales 
where they take the form of a specific new (quantum) "dark interaction" term. 
Moreover, they manifest themselves in terms of a huge number of extremely soft collective excitations.

Our result for $r_0$ confirms previous determination of the scale $r_0$ found matching 
the Newtonian acceleration of  baryonic matter with  the cosmological acceleration, 
$a_B=H$~\cite{Verlinde:2016toy,Cadoni:2017evg,Cadoni:2018dnd}. 
For a typical spiral galaxy like the Milky way, with $m_B=10^{11}M_{\odot}$, $r_0\sim Kpc$ 
which is of the order of magnitude of the distance at which one observes deviations 
from Newtons law in galaxies~\cite{Cadoni:2017evg}.
\medskip

Let us conclude this section by showing that the ground state energy (\ref{K2}) can 
be also found by using a very rough, potential well, approximation for the potential. 
 In this case, $V(x)$ is taken to be zero inside the sphere of radius $r$ and to 
jump to the asymptotic value $|V_N(L)|$ outside the sphere.
We have therefore a potential well of height $V_0$, where $V(x)=0$ for $-r<x<r$ 
and $V(x)=V_0$ for $x<-r$ and $x>r$, with $V_0=V_N(L)= m^*|\phi(L)|=\frac{G m^* m_B}{L}$.

The energy spectrum $\varepsilon$ can be found solving the Schr\"odinger 
equation. Considering   states of even parity the solution 
of the Schr\"odinger equation gives, for $\varepsilon<V_0$, the bound states,
\be\label{se}
\sqrt{2m^*(V_0-\varepsilon)}= \sqrt{2m^* \varepsilon}\tan \biggl(\frac{r\sqrt{2m^*\varepsilon}}{\hbar}\biggr).
\ee
For very small energy $\varepsilon$ ($\varepsilon\ll V_0,\, r\sqrt{2m^*\varepsilon}\ll \hbar$) 
and at leading order Eq.~(\ref{se}) gives for the ground state the energy $\varepsilon=\sqrt{\frac{V_0}{2 m^*}}\frac{\hbar}{r}$, 
which exactly matches with Eq.~(\ref{K2}).
%

%%%%%%%%%%%%%%%%%%%%%%%%%%%%%%%%%%%%%%%%%%%%%%%%%%%%
%%%%%%%%%%%%%%%%%%%%%%%%%%%%%%%%%%%%%%%%%%%%%%%%%%%%
%					ROTATIONAL CURVES
%%%%%%%%%%%%%%%%%%%%%%%%%%%%%%%%%%%%%%%%%%%%%%%%%%%%
%%%%%%%%%%%%%%%%%%%%%%%%%%%%%%%%%%%%%%%%%%%%%%%%%%%%

\section{Macroscopic effects of long-range quantum gravity at galactic scales}
\lb{sect:rot}
In the previous sections, we have argued about the presence at galactic scales of a 
(dark) interaction term (\ref{nf}) between baryonic matter and DE, which is generated 
by long-range quantum gravity effects in the DE condensate. 
We have also shown that this implies the presence of a huge number of extremely soft 
DE bosonic excitations with energy $\varepsilon\ll \hbar/L$ and wavelength $\lambda\gg L$.  
In this section we discuss the macroscopic effects due to these quantum excitations. 

By analogy with other known interactions, the macroscopic, classical, manifestation of this
large number of quanta should be a new dark force (DF) term at galactic scales. 
This possibility is particularly appealing because it could give an explanation of observed 
galactic dynamics without assuming the presence of dark matter. 
However, in our case the problem is much more involved than, say, the electromagnetic 
case, for which we have first formulated a classical theory of forces and then quantized it 
and found the associated quanta.

We do not have any classical theory of gravitational interaction "dressed" by the presence 
of DE and we are not able to quantize classical dS spacetime. 
For this reason,  the collective behaviour of these new quanta can  only be studied by means 
of statistical mechanics and thermodynamics. Indeed the analogy with condensed matter systems 
and the macroscopic properties of DE and baryons discussed in the previous sections 
allow us to understand the physics behind the macroscopic behaviour of the DE bosonic excitations. 
This analogy will help us to determine the macroscopic manifestation of long-range 
quantum gravity effects as a new component in the  acceleration felt by stars in galaxies and commonly 
attributed to dark matter. This acceleration can be thought as a dark acceleration or, equivalently, to a dark 
force term so we will refer to this new acceleration term as $a_{DF}$. In the next section we will use a 
generalized thermal equivalence principle to derive $a_{DF}$.

In our point-like approximation baryonic matter in the galactic core is modelled by a point with mass 
$m_B$ and the  test mass $m$ is located at distance $r$ from the 
baryonic mass $m_B$. 

Generically, we expect $a_{DF}$ to be given in terms of the number $N$ and 
the energy $\varepsilon$ of DE excitations.
In a thermodynamical, quantum mechanical picture, the DF acceleration 
$a_{DF}$ can be thought as generated by the
pressure $P$ of the gas of DE bosonic excitations in the sphere of 
radius $r$. Thus, we can write the acceleration for unit mass as 
$a_{DF}\sim P V/r \sim  PV \varepsilon$,  where $V$ is the volume 
of the sphere and we have used the fact that $\varepsilon$ scales as $1/r$. 
At galactic scales, $r\ll L$, the thermal contribution of  bosonic excitations, $TS$, 
to the internal energy $U$ is negligible and the variation of $U$ is produced by the work 
done by the pressure $P$ of the system.  
Usual extensive thermodynamics then implies $PV\sim N(\varepsilon)\varepsilon$, 
so that we can write
\be\label{11}
a_{DF}(\varepsilon)= C N(\varepsilon)\varepsilon^2 ,
\ee   
where $C$ is a constant with dimensions of (energy)$^{-2}$(lenght)$^{-1}$,  whose 
value will be determined shortly.  
Note that owing to its  elastic origin the dark forse is {\sl attractive}, hence the 
acceleration $a_{DF}$ is negative.
In the following, for simplicity, we will only consider the absolute value 
of forces and accelerations. 

The thermal contribution $TS$ to the internal energy becomes comparable  to 
the pressure term for $r_0\ll r\leq L$ and becomes dominant for $r= L$, i.e.~when the 
thermodynamical behaviour of the system is  completely dominated by the quantum 
properties of the dS horizon. Nevertheless, one can easily realize that the relation~(\ref{11}) 
holds also in this regime. In fact, we have $TS\sim N/L\sim N\varepsilon$ and the 
thermal contribution to the acceleration scales as in Eq.~(\ref{11}).
It is interesting to notice that, at least in our thermodynamical 
analogy, the long-range quantum mechanical contribution can be seen as 
an extensive volume term.  Conversely, the short-range thermal contribution 
associated to the dS horizon is an area term typical of standard black hole 
 (and dS) thermodynamics. This distinction will become  clearer in 
section~\ref{sect7}.

The extensive scaling behaviour $V\sim  N,a\sim  N$ of our 
gas of DE bosonic excitations is perfectly consistent with 
its origin from the constant energy density characterizing 
the DE. This is the extensive counterpart the of sub-extensive 
behaviour 
\be\lb{nnn}
a_B\sim \sqrt N  \varepsilon^2,
\ee
found for the Newtonian term $a_B$ in the radial acceleration~\cite{Mueck:2013mha,Cadoni:2017evg,Cadoni:2018dnd} 
(we will give more details about this point in the next Section).

%%%
%%%
In order to calculate the value of the constant $C$ in Eq.~(\ref{11}) we 
consider the limit in which the number of soft DE bosonic excitations 
becomes very large, $N \gg 1$, i.e.~when $r_0\lesssim r<L$. 
In this limit, $2\pi L\varepsilon\to 0$ and, at leading order in $2\pi L\varepsilon$, 
we have $N(\varepsilon_{DF})= (2\pi L \varepsilon)^{-1}$. 
In the same limit the universe becomes DE dominated, being the 
contribution of baryonic matter negligible. 
The behaviour of DE bosonic excitations should match that of the 
dS universe, i.e.~$\varepsilon = \hbar/L$ and $a_{DF}$ must become 
the cosmological acceleration $a_{DF}= H=1/L$~\cite{Cadoni:2017evg,Cadoni:2018dnd}.
This requirement determines the constant $C$ in Eq.~(\ref{11}) to be 
\be\lb{pop} 
C= \frac{2\pi }{\hbar^2}L.
\ee
From a pure quantum mechanical perspective this (singular) limit 
procedure corresponds to the transition from an excited system to the 
critical phase of a Bose-Einstein condensate, describing the dS universe, 
where all the excitations are in the fundamental state.  
However, since we do not fully understand the quantum nature of this condensate, 
we are not able to describe this phase transition.

Putting all together, i.e.~using Eqs.~(\ref{ht}),~(\ref{N_BE}),~(\ref{oi}) and~(\ref{pop}) 
into Eq.~(\ref{11}) we find the searched expression for the DF acceleration 
\be\label{232}
a_{DF}=  \frac{ 2\pi a_B}{e^{2\pi\sqrt\frac{ a_B}{H}}-1},
\ee
where $a_B= Gm_B/r^2$ is the Newtonian acceleration experienced by the test mass.
It is important to stress that our result does not explicitly depend on the quantum 
properties of the dS condensate. This is due to a nice cancellation of $\hbar$ in Eq.~(\ref{11}).
This is an important check of the validity of our result. 
Indeed, being a macroscopic effect at galactic scales, $a_{DF}$ must survive 
in the $\hbar \to0$ limit.

%%%%%%%%%%%%%%%%%%%%%%%%%%%%%%%%%%%%%%%%%%%%%%%%%%%%

\subsection{Comparison with observations}\label{subsect:rotcurves}
Let us now compare our result~(\ref{232}) with direct observations of galactic dynamics.
One of the most striking features of galactic dynamics is the so-called baryonic 
Tully-Fisher relation~\cite{Tully:1977fu,McGaugh:2000sr}. 
Astrophysical observations indicate that the total radial acceleration experienced 
by stars in galaxies can be split in two components, the Newtonian contribution $a_B$ 
due to purely baryonic matter and an additional (dark force) term $a_{DF}$: $a^r= a_B+a_{DF}$.
A simple explanation of the Tully-Fisher relation has been given by Milgrom's MOdified 
Newtonian Dynamics~(MOND)~\cite{Milgrom:1983ca,Milgrom:2014usa} in which no 
dark matter is present  in the model, whereas the additional acceleration component 
is given by $a_{DF}=a_{\rm MOND}=\sqrt{a_0a_{\rm B}}$ and $a_0$ is considered as a 
fundamental constant of nature. 
Moreover, from astrophysical observations, we know that the MOND component becomes 
relevant when $a_B\simeq a_0$ to which corresponds a critical scale $r_0\simeq \sqrt{Gm_B/a_0}$.

The observational data about rotational curves of galaxies can be fully explained by 
introducing a phenomenological interpolating function $F(x),\ x=a_B/a_0$, such that 
the total radial acceleration can be written as $a^r= F(x) a_B$~\cite{Milgrom:1983ca,McGaugh:2008nc,Sanders:2014xta}.
For $x\gg1$, near to the galactic core, the function $F$ must reproduce 
standard Newtonian gravity. 
Instead, for $x\ll1$ we have the MOND regime, $F(x)\simeq \sqrt{\frac{1}{x}}$ 
and the radial acceleration is $a^r=a_{DF}=\sqrt{a_Ba_0} =\sqrt{a_0 Gm_B/r^2}$.

An interpolating function which satisfies all the conditions above 
and fits a large amount of observational data coming from galaxies with 
different shapes (spiral, elliptical, spherical) has been proposed by 
McGaugh {\sl et al.}~\cite{McGaugh:2016leg,Lelli:2017vgz},~i.e. 
$F(x)= \frac{1}{1- e^{-\sqrt x}}$.
They measure $F(x)$ in a survey of rotation curves of 153 galaxies 
in the SPARC database. They measure $a^r$ at 2693 radii on 
these rotation curves and, at the same radii, they estimate the 
Newtonian gravitational potential from baryons as observed in 
stars, gas and dust, and so determine $a_B$. 

The McGaugh form for $F(x)$ leads to the additional acceleration term
\be\label{23}
a_{DF}= \frac{a_B}{e^{\sqrt\frac{a_B}{a_0}}-1}.
\ee
In Eq.~\eqref{23}, $a_0$ is a fitting parameter. The value of $a_0$ found by 
McGaugh {\sl et al.} corresponds, approximatively, to $a_0=H/2\pi$~\cite{McGaugh:2016leg}. 
As noted in~\cite{McGaugh:2016leg,Mannheim:2019aap} no dark matter 
is needed to fit the data since, from Eq.~\eqref{23}, it clearly appears that the 
dynamics of stars in galaxies only depends on the total amount of baryonic matter 
in the galactic region. 

Eq. \eqref{23} reproduces  the Newtonian regime of gravity 
for $a_B/a_0\to \infty$, implying $a_{DF}=0$ and $a^r= a_B$, 
whereas the  deep MOND regime, $a_{DF}=\sqrt{a_Ba_0} =\sqrt{a_0 Gm_B/r^2}$, 
can be obtained as the limiting case $a_B/a_0\to 0$.

By comparing Eq.~(\ref{23}) with Eq.~(\ref{232}) we see that our prediction 
reproduces the phenomenological result in~\cite{McGaugh:2016leg} up to a 
numerical factor, $1/(2\pi)$, which appears as a proportionality factor between 
the MOND parameter $a_0$ and the cosmological acceleration $H$.  
Indeed, our computation based on scaling arguments and 
on a simplified quantum one-dimensional model has some intrinsic limitations.
In Ref.~\cite{Tuveri:2019uej} we have proposed a general principle, namely the equality 
between the baryonic acceleration and the response of the DE medium, to 
determine this factor and reproduce in this way the phenomenological acceleration 
of McGaugh {\sl et al.}~\cite{McGaugh:2016leg,Lelli:2017vgz}. 
In the next section we will derive this proportionality factor using a generalized 
(thermal) equivalence principle.

The Newtonian regime of gravity $a_B/a_0\gg 1$ correspond to hard DE bosonic 
excitations with $\varepsilon/T_{dS}\gg1$. In this regime $N$ goes to zero exponentially 
and the DF acceleration is switched off, i.e.~$a_{DF}=0$. Conversely, the MOND regime 
$a_B/a_0\to 0$ corresponds to an huge number $N\gg1$ of  extremely soft DE bosonic 
excitations with $\varepsilon/T_{dS}\ll 1$. 
The MOND acceleration is therefore the macroscopic manifestation of an huge number 
of extremely soft long range quantum excitation of the dS universe generated by the 
presence of baryonic matter. 

%%%%%%%%%%%%%%%%%%%%%%%%%%%%%%%%%%%%%%%%%%%%%%%%%%%%%%%%%%%%%%%%%%%%%%
%%%%%%%%%%%%%%%%%%%%%%%%%%%%%%%%%%%%%%%%%%%%%%%%%%%%%%%%%%%%%%%%%%%%%%

%				A generalized  thermal equivalence principle

%%%%%%%%%%%%%%%%%%%%%%%%%%%%%%%%%%%%%%%%%%%%%%%%%%%%%%%%%%%%%%%%%%%%%%
%%%%%%%%%%%%%%%%%%%%%%%%%%%%%%%%%%%%%%%%%%%%%%%%%%%%%%%%%%%%%%%%%%%%%%

\section{A generalized thermal equivalence principle}
 \lb{sec:4}
In a recent paper Smolin has proposed that the MOND theory corresponds to a quantum 
gravity, cosmological constant-dominated, regime in which the equivalence principle does 
not hold in its usual form~\cite{Smolin:2017kkb}.  
This is the same quantum regime of temperatures below the dS temperature and with length 
scales larger than the cosmological horizon we are considering in this paper. 
Similarly to our description, the MOND acceleration is seen as a macroscopic long-range 
quantum gravity effect, which survives in the $\hbar \to 0$ limit. 
Note also that an extension of \cite{Smolin:2017kkb} which points in a direction similar to 
what we have discussed in the previous sections has been done in Ref.~\cite{Alexander:2018lno}, 
where dark matter effects at galactic scales are described by means of superfluids particles. 

Although starting from the same general idea, namely the existence of a long-range, quantum gravity, 
cosmological constant-dominated regime, our approach differs from that of Ref.~\cite{Smolin:2017kkb}. 
Smolin explains the modification of Newtonian gravity laws for $a_B<a_0$ as originated by an 
environmental-dependent change of the ratio $m_i/m_g$  between the gravitational and inertial 
mass, $m_g$ and $m_i$, respectively. Conversely, in our approach the modification of Newtonian 
gravity is a direct effect of long-range quantum DE excitations.

It is quite clear that the crucial question for both approaches concerns the fate of the classical equivalence 
principle in the region $a_B\ll a_0$, when long-range quantum gravity effects become dominant: 
does a quantum extension of Einstein's equivalence principle exist?

Actually, as remarked in Ref.~\cite{Smolin:2017kkb} there are two formulations of the equivalence principle, 
which are classically equivalent but which differ when we consider excitations of wavelength of the 
same order of magnitude of the curvature radius of the spacetime $\cal{R}$ (in our case $L$). 
 This, in turns, is what we are discussing in this paper.
The first formulation (called EP1 in Ref.~\cite{Smolin:2017kkb}) asserts that locally, i.e. for observers 
with extent $l\ll \cal{R}$, a gravitational field can be eliminated by free fall.  
The second formulation (called EP2 in Ref.~\cite{Smolin:2017kkb}) asserts that to zeroth order in 
$l/\cal{R}$ gravity can be mimicked by uniformly accelerated observers. 

The quantum version of EP2 proposed in Ref.~\cite{Smolin:2017kkb} takes the form of a 
{\bf thermal equivalence principle (TEP)} which incorporates universality of free fall by asserting that 
{\sl the temperature $T$ seen by an observer is related to  his acceleration $a$ by the Deser-Levin 
(DL) formula} \footnote{To make clear the nature of the effects we are considering in this section we 
will use conventional units, i.e. units in which the speed of light $c$ is not set to $1$. }
\be\lb{dl}
T=\sqrt{T^2_{dS}+ \left(\frac{\hbar a}{2\pi c}\right)^2}
\ee
where $T_{dS}$ is the dS temperature as given in Eq. (\ref{ht}).

An extension of EP1 to the quantum, $\Lambda$-dominated, regime is excluded in Ref.~\cite{Smolin:2017kkb} 
with the argument that in this regime the relevant phenomena are not small compared to $\cal{R}$ 
(we are considering excitation modes with wavelength much bigger than $L$!) thus making them 
intrinsically not compatible with the restrictions required by EP1.
However, looking for a quantum generalization of the classical EP1, the restriction that $l\ll \cal{R}$ seems 
too strong, particularly in view of the non-local character of quantum effects. Moreover, the TEP as 
formulated in Ref.~\cite{Smolin:2017kkb}  shows an intrinsic quantum/classical asymmetry: 
it predicts the temperature of quantum modes generated by a macroscopic and classical cause 
(the acceleration) but says nothing about the macroscopic effect of these quantum modes.

We propose to generalize the TEP of Ref.~\cite{Smolin:2017kkb} by formulating it in a symmetric 
way  in the sense expressed above.
The {\bf generalized thermal equivalence principle (GTEP)} asserts not only the statement 
quoted above  in the TEP formulation of Smolin but also that {\sl whenever we have a 
thermal ensemble at temperature $T$ of quantum gravity degrees of freedom, the macroscopic 
acceleration produced on a test mass is given by the Deser-Levin formula (\ref{dl})}, 
\be\lb{GTEP}
a=\frac{2\pi c}{\hslash}\sqrt{T^2-T^2_{dS}}.
\ee 
Let us now show that our GTEP predicts correctly the behaviour of black holes and the dS 
universe considered as quantum systems. Moreover, we will show that in the case of 
long-range quantum gravity excitations the GTEP implies our formula~(\ref{11}) and 
allows us to derive our Eq.~(\ref{232}) with a proportionality factor between 
$a_0$ and $H$ consistent with observations.
 
%%%%%%%%%%%%%%%%%%%%%%%%%%%%%%%%%%%%%%%%%%%%%%%%%%%%%%%%%%%%%%%%%%
 
\subsection{Black holes}
Astrophysical black holes correspond to the limiting case $T\gg T_{dS}$ of Eq. (\ref{GTEP}),
\be\lb{dl1}
a=\frac{2\pi c}{\hbar }T,
\ee
which, as expected, is nothing but the relationship between surface gravity and the 
Hawking temperature for a Schwarzschild black hole. For a black hole of mass $M$ 
and Schwarzschild radius $R_s=2GM/c^2$ this relation can be also written as 
$a=\frac{ c^2}{2 R_s}$.
In the most simple-minded quantum model of a black hole, we can consider it as 
an ensemble of quanta with typical energy
\be\lb{bhe} 
\varepsilon= \frac{\hbar c}{R_s}.
\ee
Using the previous equation into Eq.~(\ref{dl1}) we find the nice relation between 
the energy of the quanta and the acceleration
\be\lb{dl2}
a=\frac{c}{2\hbar}\varepsilon. 
\ee
The same result can be obtained in the corpuscular model of black holes  of 
 Refs~\cite{Dvali:2013eja,Dvali:2012rt,Casadio:2016zpl,Cadoni:2017evg,Cadoni:2018dnd}. 
Staring from the  acceleration in Eq.~(\ref{nnn}), where the number $N$ of gravitons is 
related to the black hole mass by $N\sim M^2/m_p^2$ and by using Eq.~\eqref{bhe}, we find 
Eq.~(\ref{dl2}) apart from a dimensional proportionality constant to be chosen appropriately. 
 
%%%%%%%%%%%%%%%%%%%%%%%%%%%%%%%%%%%%%%%%%%%%%%%%%%%%%%%%%%%%%%%%%%%% 
 
\subsection{The de Sitter universe}
 Similarly to black holes, the GTEP also applies to dS universe. 
In this case, Eq. (\ref{GTEP}) takes the form $a=\frac{2\pi c}{\hbar }T_{dS}$, i.e  we find the same equation
 as in~(\ref{dl1}) valid for black holes, 
with $T_{dS}$ replacing the black holes temperature $T$. 
For the dS universe, Eq.~(\ref{dl1}) gives the well-known relation between 
the dS temperature and the cosmological acceleration $H$~\cite{Narnhofer:1996zk,Deser:1997ri}:
\be\lb{ds2}
a=H=\frac{2\pi c}{\hbar }T_{dS}=\frac{c^2}{ L }.
\ee
 Hence, the dS universe can be thought as an ensemble of quanta with typical energy
\be\lb{bhe1} 
\varepsilon= \frac{\hbar c}{L},
\ee
so that we find 
\be\lb{dl4}
a=\frac{c}{\hbar}\varepsilon,
\ee
which  is similar to Eq. (\ref{dl2}) up to a factor of $2$.  

Also in this case the  corpuscolar picture of gravity offers a good setup to find Eq.~\eqref{dl4}~\cite{Cadoni:2018dnd,Binetruy:2012kx}.
Here too the acceleration is given by Eq.~(\ref{nnn}), where the number of gravitons $N$ 
is determined by the dS radius as $N\sim L^2/l_p^2$. Using this equation together with Eq.~(\ref{bhe1}) 
into Eq.~(\ref{nnn}) we find, choosing appropriately the proportionality constant, Eq.~(\ref{dl4}).

%%%%%%%%%%%%%%%%%%%%%%%%%%%%%%%%%%%%%%%%%%%%%%%%%%%%%%%%%%%%%%%%%%%

\subsection{Long-range quantum gravity regime}
Let us now apply the GTEP to the long-range quantum gravity regime proposed 
in this paper, i.e.~to DE bosonic excitations generating the DF.
The regime we are considering now is that of a thermal ensemble of DE bosonic 
excitations with temperature $T$ very close to $T_{dS}$. 
The DL formula~\eqref{GTEP} holds only for $T>T_{dS}$ thus we cannot describe 
the thermal excitations with $T\le T_{dS}$. This is consistent with the fact that at $T\simeq T_{dS}$ 
the system undergoes a phase transition.

 However, let us expand the DL formula~\eqref{GTEP} near a temperature $T_1= \sigma T_{dS}$ 
where $\sigma>1$ and is of order $1$. We find, at leading order in $T^2$,
\be\lb{jjj}
a= \rho H+ \eta \frac{L}{\hbar^2}T^2,
\ee
where $H$ is the cosmological acceleration and $\rho,\eta$ are dimensionless constants, which, 
as expected, blow up for $\sigma=1$. The first term describes the background cosmological dS 
acceleration, whereas the second one describes  the thermal bath of DE bosonic excitations. 
Taking into account that  the typical energy of an excitation of the thermal bath is $\varepsilon\sim T $ the second 
term in Eq.~(\ref{jjj}) can be written in the following form,
\be\lb{kkk}
a\sim \eta\frac{L}{\hbar^2}\varepsilon^2.
\ee
This formula is very similar to Eq.~(\ref{11}), which gives  the acceleration produced by a 
number $N$ of bosonic DE excitations of energy $\varepsilon$. In order to have a complete 
matching between Eq.~\eqref{kkk} and Eq.~\eqref{11},  we have to consider in Eq.~(\ref{11}) 
the contribution of a single quantum mode ($N(\varepsilon)\sim1$) and to fix the proportionality 
constant $\eta$ in Eq. ~\eqref{kkk} to the same value $2\pi/\hbar^2 $ given by Eq.~(\ref{pop}).
We get the significant result
\be\lb{kkkk}
a=\frac{2\pi L}{\hbar^2}\varepsilon^2.
\ee
Let us stress again  that this results only holds in the regime in which the temperature $T$ is  close to $T_{dS}$. 
\medskip

 A comparison of this result with those derived for black holes in Eq.~(\ref{dl2}) and for the 
dS universe in Eq.~(\ref{dl4}) can be very instructive.
All the three equations link the gravitational acceleration to the energy of the quanta and are 
a direct consequence of the GTEP applied in different quantum gravity regimes. 
The black hole and the dS cases in Eq.~(\ref{dl2}) and Eq.~(\ref{dl4}), respectively, 
although corresponding to completely different length scales regimes, share the 
same result: the gravitational acceleration scales linearly with the energy of the quanta. 
This is a {\sl quantum relativistic effect} because it goes to zero when the speed of light $c$ 
goes to zero.
Conversely, in the case of DE bosonic excitations~(\ref{kkkk}), the acceleration scales 
quadratically with the energy of the quanta. Moreover, being independent of the speed of 
light, the effect survives in the $c\to 0$ limit. This is perfectly consistent with the  
{\sl non-relativistic nature of the MOND theory}. 

In section~\ref{sect:rot} we have  derived the McGaugh {\sl et al.} phenomenological 
formula~(\ref{23}) up to the proportionality factor between the MOND and the 
cosmological acceleration, $a_0$ and $H$, respectively.
This uncertainty was inherited from the approximations we have used  quantum model used in sect.~\ref{sect2} 
to determine the energy of DE bosonic excitations, $\varepsilon$.  

 Let us now show how to recover the exact proportionality factor ($1/2\pi$) in the context 
of the GTEP.  
For a single quantum state, the gravitational acceleration in Eq.~\eqref{kkkk} 
becomes equal to the gravitational acceleration generated by the baryonic mass $m_B$, 
i.e.~$a=a_{DF}=a_B$.
Using this equation in Eq. (\ref{kkkk}) we get
\be\lb{ddd}
\varepsilon=\sqrt{\frac{Gm_B}{2\pi L}}\frac{\hbar}{r}.
\ee
This allows us to compute the DF acceleration felt by stars in galaxies. Using 
Eqs.~\eqref{ddd},\eqref{N_BE} and ~\eqref{pop} into  Eq.~\eqref{11} we get 

\be\label{2321}
a_{DF}=  \frac{a_B}{e^{\sqrt\frac{ 2 \pi\, a_B}{H}}-1},
\ee
which is exactly the McGaugh {\sl et al.} expression (\ref{23}), with  $a_0=H/2\pi$ 
and improves  Eq.~(\ref{232}) by determining the  proportionality factor between $a_0$ and $H$.
Note that the equation $a_{DF}=a_B$ has been used in Ref. \cite{Tuveri:2019uej} as a 
general principle to determine the energy of the DE bosonic excitations.

%
%%%%%%%%%%%%%%%%%%%%%%%%%%%%%%%%%%%%%%%%%%%%%%%%%%%%%
%%%%%%%%%%%%%%%%%%%%%%%%%%%%%%%%%%%%%%%%%%%%%%%%%%%%%
%     						CORPUSCOLAR GRAVITY 
%%%%%%%%%%%%%%%%%%%%%%%%%%%%%%%%%%%%%%%%%%%%%%%%%%%%%
%%%%%%%%%%%%%%%%%%%%%%%%%%%%%%%%%%%%%%%%%%%%%%%%%%%%%

%%%%%%%%%%%%%%%%%%%%%%%%%%%%%%%%%%%%%%%%%%%%%%%%%%%%%%%%%%%%%%%%%%%%%%%

\section{Galactic dynamics in a corpuscular gravity picture}
\lb{sect7}
In the previous sections, we have derived the phenomenological expression~(\ref{23}) using 
general features of quantum and statistical mechanics, thermodynamics, general relativity and 
a (thermal-)quantum generalization of the equivalence principle.

A drawback of our approach is that we do not have a detailed description of the microphysics 
involved in the gravitational interactions between baryonic matter and the DE medium which, 
in turns produces the long-range bosonic excitations discussed in the previous sections.

In order to have a more definite picture of the microphysics involved, we will work out Eq.~(\ref{oi}) 
in the  corpuscular gravity scenario of Refs.~\cite{Verlinde:2016toy,Cadoni:2017evg,Cadoni:2018dnd}. 
The same approach has been used in~\cite{Cadoni:2017evg,Cadoni:2018dnd} to determine the 
asymptotic behaviour of $a_{DF}$ in the deep MOND regime. 

 An important feature of the corpuscular gravity picture of Refs.~\cite{Cadoni:2017evg,Cadoni:2018dnd} 
is the use of Verlinde's idea~\cite{Verlinde:2016toy} about the transition between the 
Newtonian and the DE-dominated regimes of gravity as a competition between area and volume 
contributions to the number of degrees of freedom (entropy in Verlinde's paper) of the two systems. 
This is a crucial point to determine the "dark matter effects" as a reaction of DE to the presence of  baryonic matter.
Moreover, the dS universe is described as a BEC of gravitons and, similarly, the interaction 
between baryonic matter and DE is mediated by gravitons called "dark gravitons".
Differently from what we have discussed in the previous sections, in the corpuscolar gravity approach 
both the graviton number as well as the gravitons energy depend  on the size of the condensate~\cite{Dvali:2010bf,Dvali:2010ue,
Dvali:2010jz,Dvali:2011th,Dvali:2011aa,Dvali:2012en,Dvali:2012rt,Binetruy:2012kx,Dvali:2013eja}.
We recall here the typical scaling of the number of gravitons for DE, $N_{DE}$, and for baryonic 
matter $N_{B}$ in terms of the size $r$ of the system and of the baryonic mass:
\be\label{sca5}
N_{DE}(r)\sim \frac{r^3}{l_p^2L},\quad N_B(r)\sim \frac{r^2}{l_p^2},\quad N_{B}(m_B)\sim \frac{m^2_B}{m_p^2}. 
\ee
The proportionality factor for $N_{DE}(r)$ is chosen in such a way that it matches with 
the total number $N_{dS}= L^2/l_p^2$ of DE gravitons in the dS spacetime at $r=L$. 
The number of gravitons associated to baryonic matter and dS universe scales 
holographically.
There is a nice correspondence between the scaling of the graviton number on 
the one side and the acceleration formulae on the other side. Different scalings 
for the graviton numbers and different acceleration formulae characterize different 
regimes of the gravitational interaction.
Eq. (\ref{sca5}) tells us that the graviton number $N_{DE}$ associated with   DE  has a  volume contribution 
which becomes relevant at intermediate scales. 
Translated in the language of the previous sections, this means that the long-range 
quantum mechanical effects of DE generated by the presence of baryonic matter 
are responsible for the extensive scaling $N_{DE}$ at intermediate scales shown in 
Eq.~\eqref{sca5}. 
In the corpuscular picture, critical gravitational systems as black holes and dS universe 
can be described  in terms of the scaling of the   gravitons number  $N$, which 
behaves holographically. 

This has a counterpart in  the non-extensive behaviour of Eq.~(\ref{nnn}) for the gravitational 
acceleration and in  its linear form as a function of the energy of the gravitons, 
$\varepsilon$ given in Eqs.~(\ref{dl2}),(\ref{dl4}). 
Conversely, the extensive behaviour for the DE gravitons' number is associated to 
non-critical gravitational systems (the long-range quantum gravity regime). It is 
in correspondence with the extensive behaviour expressed in Eq.~(\ref{11}) for the 
acceleration and with its quadratic expression in terms of $\varepsilon$ given in Eq.~(\ref{kkkk}).

In the Newtonian regime, the universe is dominated by baryonic matter, $N_{B}> N_{DE}$. 
The Newtonian potential can be described by means of a quantum coherent state of 
$N_B$ gravitons with Compton length $r$,~i.e. of energy~\cite{Cadoni:2018dnd,Casadio:2016zpl}  
\be\label{scal}
 \varepsilon_{N}\simeq\frac{\hbar}{r}.
 \ee
The transition from the Newtonian to the MOND regime of gravity occurs when $N_B(r)=N_{DE}(r)$. 
This determines the scale $r_0$ at which the dark force effects become relevant~\cite{Verlinde:2016toy,Cadoni:2018dnd} 
 and it is exactly given by Eq.~(\ref{ppp}).
Following Ref.~\cite{Cadoni:2018dnd} (see also~\cite{Verlinde:2016toy}) the effect of baryonic matter 
on the DE condensate is the subtraction of DE gravitons according to the formula
\be\label{conserv}
 \delta N_{DE}=-N_B.
 \ee
The dark force is  then interpreted as  the response of the DE condensate to the presence of baryonic 
matter so that the energy of the gravitons mediating the interaction will be modified as follows
\be\label{ener}
  \varepsilon_{N}+ \delta \varepsilon_{N}=\varepsilon_{N}+\varepsilon_{DF}
\simeq \frac{\hbar}{r}- \frac{\hbar}{r} \frac{\delta r}{r} \ee
The first terms describes the Newtonian interaction whereas the second term represents the energy 
of the gravitons producing the dark force. Because it is relevant only at scales of order $r_0$, the latter 
can be written as
\be\label{ener}
 \varepsilon_{DF}= - \frac{\hbar \alpha}{r},\quad \alpha= \frac{\delta r}{r}\Big|_{r=r_0}.
 \ee
The term $\delta r|_{r=r_0}$ can be computed using Eqs.~(\ref{sca5}) into Eq.~(\ref{conserv}) giving 
$\delta r|_{r=r_0}\simeq- Gm_B$.  It is interesting to notice that $\delta r$ evaluated at the scale $r_0$ 
has the same sign and the same order of magnitude of $\delta r$ evaluated at the scale of the cosmological 
horizon: $\delta L=- 2Gm_B$~\cite{Cadoni:2018dnd,Verlinde:2016toy}. 
Physically this means that baryonic matter subtracts a certain amount of gravitons (entropy in \cite{Verlinde:2016toy}) 
from  the DE condensate.
Using this result and Eq.~(\ref{ppp}) one easily obtains 
\be\label{ener2}
 \alpha\simeq \sqrt{\frac{ Gm_B}{L}},
 \ee
which as expected reproduces Eq.~(\ref{oi}).

It is quite interesting to notice that the expression for the DF coupling constant $\alpha$ we have found 
can be rewritten in terms of  a simple ratio between the baryonic mass $m_B$ and the total DE mass, 
$m_{DE}$  or, equivalently, between the baryonic gravitons' number $N_B$ and the total number of DE 
gravitons in the dS spacetime, $N_{dS}$. Indeed, exploiting the quadratic scalings of the masses and graviton 
number given in Eq.~(\ref{scal}), we can rewrite Eq.~(\ref{ener2}) as follows
\be\label{pp}
\alpha\simeq \sqrt{\frac{m_B}{m_{dS}}}\simeq \left(\frac{N_B}{N_{dS}}\right)^{1/4},
\ee
where $m_{dS}$ is the total mass of the dS universe. Because dark force gravitons are pulled out from the DE 
condensate by the baryonic matter, Eq.~(\ref{pp}) tells us that the DF coupling constant is determined completely 
by the fraction of gravitons of the DE condensate subtracted by baryonic matter.
%
%Using this expression of $\alpha$ to compute $\varepsilon$ in 
%Eq.~(\ref{nf}) and then comparing Eq.~(\ref{19a} with Eq.~(\ref{23}), 
%one finds the parameter $\gamma$ of Eq.~(\ref{pp})
%%
%\be\lb{de}
%\gamma= \frac{\pi^2 \sigma^2} {18 \beta^2}.
%\ee
%%
%For critical matter (e.g. black holes and the DEC) the relation 
%$r=2 G m_B$ holds which, in turns, implies  $\sigma/\beta= 4$ $ \gamma=(8/9) \pi^2$. 
%This is somehow bigger then the value of $\gamma$ for which Eq.~(\ref{19a}) 
%exactly matches Eq.~(\ref{23}). This value is $\gamma=2\pi$ giving
%%
%\be\lb{de3}
%\frac{\sigma}{\beta}= \frac{6}{\pi}.
%\ee
%
%Non critical baryonic matter could produce  a smaller value  of $\gamma$  however.....

%%%%%%%%%%%%%%%%%%%%%%%%%%%%%%%%%%%%%%%%%%%%%%%%%%%%%
%%%%%%%%%%%%%%%%%%%%%%%%%%%%%%%%%%%%%%%%%%%%%%%%%%%%%
%     						METRIC DESCRIPTION
%%%%%%%%%%%%%%%%%%%%%%%%%%%%%%%%%%%%%%%%%%%%%%%%%%%%%
%%%%%%%%%%%%%%%%%%%%%%%%%%%%%%%%%%%%%%%%%%%%%%%%%%%%%

%%%%%%%%%%%%%%%%%%%%%%%%%%%%%%%%%%%%%%%%%%%%%%%%%%%%%%

\section{General relativity uplifting and effective fluid description}
\lb{sec:8}
One common problem of  MOND and others  approaches where infrared modifications 
of the laws of gravity are used to describe the "dark matter phenomenology"  
is the difficulty to perform a ``metric-covariant uplifting'' of the theory~\cite{Famaey:2011kh}.
They are usually formulated in the weak-field regime, whereas gravity must allow for the metric-covariant 
description given by GR.
In our approach the gravitational interaction at galactic scales has an infrared contribution coming from 
long-range quantum effects. Therefore, fluid space-time models are the natural candidates for providing 
us with a simple way to perform a metric-covariant uplifting of our theory. We look for IR modifications of 
Einstein gravity in which the Einstein-Hilbert action remains unchanged and long-range quantum effects 
are modelled as a fluid sourcing a classical gravitational field. 

In Refs.~\cite{Cadoni:2017evg,Cadoni:2018dnd} it has been shown that the emergent laws of gravity 
based on a BEC of gravitons can be described as GR sourced by an anisotropic fluid.
In this description the radial pressure of the fluid describes the dark force.  
Since the latter is originated from the quantum long-range properties of DE, if we neglect them, the DE is 
in general described as an isotropic fluid with constant energy density and equation of state $p=-\rho$.  
The simplest way to take into account at the GR level the reaction of DE to the presence of baryonic 
matter responsible for the dark force effects is, therefore, to consider an anisotropic fluid as source in 
Einstein's field equations.

This effective fluid description can be easily generalised to the case under consideration in this paper, 
considering a GR completion of the weak-field description given by Eq.~(\ref{23}). 
Following Refs.~\cite{Cadoni:2017evg,Cadoni:2018dnd}, 
the total radial acceleration has a Newtonian and a pressure term 
\be\label{1}
a^r=a_B+ 4\pi Gr \ppar.
\ee
The pressure profile $\ppar$ has to be chosen in such way that Eq.~(\ref{1}) matches Eqs.~\eqref{23}:
\be\label{4}
\ppar(r)= \frac{1}{4\pi}\frac{m_B}{r^3}\frac{1}{e^{\frac{1}{r}\sqrt{Gm_B/a_0} }-1}
\ee 
and the radial acceleration is 
\be\label{5}
a^r= \frac{G m_B}{r^2}\frac{1}{1-e^{\frac{1}{r}\sqrt{Gm_B/a_0} }}.
\ee
The full metric solution can be obtained solving Einstein field equations sourced by an 
anisotropic fluid with a pressure profile given by Eq.~(\ref{4}).  
The spacetime metric is taken of the form $ds^2= -f(r)e^{\Gamma(r)}dt^2+\frac{dr^2}{f(r)}+r^2d\Omega^2$ 
and it is given by
\be\label{6}
 f=1-\frac{2Gm_B}{r},\quad   
\Gamma'= f^{-1}\left(
\frac{2Gm_B}{r^2}\frac{1}{e^{\frac{1}{r}\sqrt{Gm_B/a_0} }-1}\right).
\ee
In the weak field limit of the metric one finds the potential $\phi= \frac{1}{2}(fe^\Gamma)$, 
from which one can easily derive the form of the DF 
%components for the 
acceleration and check that it exactly reproduces Eq.~(\ref{23}).
In the MOND regime of Eq.~(\ref{23}), i.e. $a_B/a_0\to 0$, one obtains the same expansion 
of the gravitational potential generated by a point-like particle given in Ref.~\cite{Cadoni:2017evg} 
with the characteristic logarithmic behaviour of MOND and an extremely tiny Machian contribution 
to the Newtonian potential~(see Ref.~\cite{Cadoni:2017evg}).

%%%%%%%%%%%%%%%%%%%%%%%%%%%%%%%%%%%%%%%%%%%%%%%
%%%%%%%%%%%%%%%%%%%%%%%%%%%%%%%%%%%%%%%%%%%%%%%
%				CONCLUSIONS
 %%%%%%%%%%%%%%%%%%%%%%%%%%%%%%%%%%%%%%%%%%%%%%%
%%%%%%%%%%%%%%%%%%%%%%%%%%%%%%%%%%%%%%%%%%%%%%%

\section{Conclusions}
\lb{sec:9}
In this paper we have argued that long-range quantum gravity effects  could play a relevant 
role at galactic scales.
The non-Newtonian behaviour of the radial acceleration at these scales is explained, without assuming 
the presence of dark matter, as the macroscopic effect of a huge number of extremely soft bosonic 
DE excitations with wavelength much bigger than the size of the cosmological horizon, in thermal 
equilibrium with de Sitter spacetime.

We have used simple and general arguments not relaying on any specific assumption of the would-be 
microscopic theory of gravity but,  rather, only on general features of thermodynamics, quantum 
and statistical mechanics and general relativity.

Our formula for the additional  "dark" component of the acceleration agrees with the  phenomenological 
relation obtained by McGaugh {\sl et al.}  by fitting a large amount of observational data and with the 
MOND theory.
The functional form of the dark acceleration is a direct consequence of the Bose-Einstein thermal 
distribution for the number of DE excitations. 

The BE distribution, in particular the exponential suppression of the number of hard modes, together 
with the expression for the energy $\varepsilon$ of these modes, explain why the number of DE modes 
becomes significant only at galactic scales, i.e.~at scales larger than the IR scale $r_0$ 
 at which the dark matter effects arise in galaxies.
For this reason, they only affect the gravitational interaction at galactic scales, leaving unaltered the 
usual Newtonian contribution at smaller ones. Moreover, the BE distribution and the explicit 
expression of $\varepsilon$ lead to the correct value of $r_0$.

Remarkably enough, we have derived the same results using a generalized thermal equivalence principle 
(GTEP), which promotes the Deser-Levin formula in Eq.~\eqref{dl} to an universal dynamical feature of thermal 
quantum gravity systems. 
Moreover, the GTEP allows us to determine the value $1/2\pi$ for the proportionality factor between the MOND 
acceleration parameter $a_0$ and the cosmological acceleration $H$. This value is in accordance with the 
results  found in~\cite{McGaugh:2016leg,Lelli:2017vgz} obtained by fitting a large amount of observational 
data.

Last but not least we have shown that our formula appears as the weak field limit of  Einstein's general relativity 
sourced by an anisotropic fluid.

A weakness of our paper is the lack of a quantum theory describing the DE bosonic excitations, which play a 
crucial role in our derivation. Presently, we do not have a consistent theory of quantum gravity of the dS spacetime, 
which,  in some way, should provide us the consistent description of these DE bosonic excitations. 
In particular, we expect that  this theory should reproduce our result (\ref{oi}). 
Similar considerations hold for the GTEP we have proposed in this paper. Owing to its intrinsic quantum origin 
we expect this principle to find an explanation in the thermal description of the quantum theory of the dS spacetime.

Our explanation of galactic phenomenology  as dominated, at large distances from the galactic core, by long-range 
quantum gravity effects presently covers only the case of galactic rotational curves.
Gravitational lensing effects are also very important for validating or confuting our approach. These effects could 
be described by using the covariant uplifting of our weak field description, namely Einstein's general relativity 
sourced by an anisotropic fluid  presented in sect. (\ref{sec:8}).  
We have not addressed this important point in this paper.

It must be also clearly stated that at present stage of development of the subject, the long-range quantum gravity 
approach we have presented here cannot represent a full alternative to dark matter and the $\Lambda$CDM model. 
All the cosmological implications of our approach, in particular those concerning structure formation  and the 
dynamical process of the bullet cluster have to be worked out if long-range quantum gravity effects have to 
represent a viable alternative to dark matter.

%%%%%%%%%%%%%%%%%%%%%%%%%%%%%%%%%%%%%%%%%%%%%%%%%%%%%%%%%%%%%%%%%%%%%%%%%%%%%%%%5

%\bibliographystyle{utphys}
%\bibliography{gravitons}
%\bibliography{gravitonsDF}
%
%\providecommand{\href}[2]{#2}\begingroup\raggedright

\providecommand{\href}[2]{#2}\begingroup\raggedright\endgroup

\end{document}